%% file: main.tex
\documentclass{article}
\usepackage{blindtext}
\usepackage[a4paper, total={6in, 10in}]{geometry}
\usepackage{amsmath,amssymb,amsfonts}
\usepackage{graphicx}
\usepackage{subcaption}
\usepackage{textcomp}
\usepackage{xcolor}
\usepackage{glossaries}
\usepackage{svg}
\usepackage{algorithm}
\usepackage[noend]{algpseudocode}
\usepackage[shortlabels]{enumitem}

\usepackage{booktabs}
\usepackage{multirow}
\usepackage[affil-it]{authblk} 
\usepackage{hyperref}

\glsdisablehyper
\input{acronyms}
\usepackage[per-mode=fraction]{siunitx} 
\usepackage{cleveref}
\usepackage{array} 
\usepackage{geometry} 
\newcolumntype{C}[1]{>{\centering\arraybackslash}m{#1}}
\usepackage[numbers, square]{natbib}

\def\BibTeX{{\rm B\kern-.05em{\sc i\kern-.025em b}\kern-.08em
    T\kern-.1667em\lower.7ex\hbox{E}\kern-.125emX}}

\AtBeginDocument{%
  \providecommand\BibTeX{{%
    \normalfont B\kern-0.5em{\scshape i\kern-0.25em b}\kern-0.8em\TeX}}}
\DeclareSIUnit[per-mode=symbol]\tps{\transaction\per\second}
\DeclareSIUnit[per-mode=symbol]\kbps{\kilo\bps}
\DeclareSIUnit[per-mode=symbol]\Mbps{\mega\bps}
\DeclareSIUnit[per-mode=symbol]\Gbps{\giga\bps}
\DeclareSIUnit[per-mode=symbol]\nanosec{\nano\second}
\DeclareSIUnit\microsec{\SIUnitSymbolMicro s}
\DeclareSIUnit\byte{B}
\DeclareSIUnit\bit{bit}
\DeclareSIUnit\transaction{transaction}
\DeclareSIUnit\terabyte{TB}

\title{Playing the MEV Game on a First-Come-First-Served Blockchain}

\author{Burak {\"{O}z}}
\author{Jonas Gebele}
\author{Parshant Singh}
\author{Filip Rezabek}
\author{Florian Matthes}

\affil{Technical University of Munich}

\makeatletter
\let\@date\relax 
\makeatother

\begin{document}
\maketitle

\begin{abstract}
Maximal Extractable Value (MEV) searching has gained prominence on the Ethereum blockchain since the surge in Decentralized Finance activities. In Ethereum, MEV extraction primarily hinges on fee payments to block proposers. However, in First-Come-First-Served (FCFS) blockchain networks, the focus shifts to latency optimizations, akin to High-Frequency Trading in Traditional Finance. This paper illustrates the dynamics of the MEV extraction game in an FCFS network, specifically Algorand. We introduce an arbitrage detection algorithm tailored to the unique time constraints of FCFS networks and assess its effectiveness. Additionally, our experiments investigate potential optimizations in Algorand's network layer to secure optimal execution positions.

Our analysis reveals that while the states of relevant trading pools are updated approximately every six blocks on median, pursuing MEV at the block state level is not viable on Algorand, as arbitrage opportunities are typically executed within the blocks they appear. Our algorithm's performance under varying time constraints underscores the importance of timing in arbitrage discovery. Furthermore, our network-level experiments identify critical transaction prioritization strategies for Algorand's FCFS network. Key among these is reducing latency in connections with relays that are well-connected to high-staked proposers.
\end{abstract}

\section{Introduction}
Blockchain networks offer profitable opportunities for various parties involved. While default economic incentives for consensus participants include block rewards and transaction fees, the recent surge in \gls{defi} activities (with daily trading volumes on \glspl{dex} exceeding multi-billion USD~\cite{top_gecko}) has led to the emergence of a new incentive for strategically acting players, which we know as \gls{mev}~\cite{daian}. While this term generally refers to the value that can be captured by entities like block proposers, who have the privilege to determine transaction inclusion, exclusion, and ordering, value extraction is not limited to them, as \gls{mev} activity dashboards such as libMEV~\cite{noauthor_libmev_nodate} reflect a total of \SI{64}{million} USD made by profit-seeking entities operating on the mempool, known as \textit{\gls{mev} searchers}, since the merge on Ethereum in September 2022. 

Before private relays such as Flashbots\footnote{Flashbots Auction: \url{https://docs.flashbots.net/flashbots-auction/overview}} emerged on Ethereum, offering an off-chain sealed-bid auction for inclusion, \gls{mev} searching bots competed in \glspl{pga} on the public mempool to extract value~\cite{daian}. As Ethereum is a blockchain where fees can influence the transaction ordering of proposers, while latency still played a role in being competitive in \glspl{pga} (e.g., to ensure your latest bid reaches the proposer on time), the primary determinant of the winner was the fee offered.

In blockchains where block proposers arrange transactions in the order received, known as \gls{fcfs}, the dynamics for \gls{mev} searching differ, as demonstrated in \cite{carrillo_mev_2023, oz_study_2023}. Unlike fee-based blockchains like Ethereum, in an \gls{fcfs} network, the only way to prioritize is by propagating a transaction before competitors. Consequently, the available runtime for an \gls{mev} searching algorithm in \gls{fcfs} networks is limited by the anticipated arrival time of a competing transaction, as opposed to almost the entire block time available to the searcher on a fee-based network.

This paper explores how the \gls{mev} game can be strategically played on an \gls{fcfs} network, using Algorand as the case study, building on the initial research by Öz et al.~\cite{oz_study_2023}. Our methodology comprises two stages: \textit{discovery} and \textit{extraction}. The discovery stage employs a cyclic arbitrage detection algorithm. We focus on arbitrages, as they are shown to be a feasible \gls{mev} strategy on \gls{fcfs} networks~\cite{carrillo_mev_2023, oz_study_2023}. We assess our algorithm's performance using historical Algorand data, considering the runtime constraints of \gls{fcfs} networks. In the next stage, similar to High-Frequency Trading in  Traditional Finance, \gls{mev} extraction in \gls{fcfs} networks is about outpacing others in latency. We run experiments on a private Algorand network to identify key factors for prioritizing transaction propagation. Our research contributes to understanding \gls{mev} in \gls{fcfs} networks, showcasing how algorithms can exploit profitable opportunities and highlighting the critical network characteristics for ensuring prioritized execution.


\section{Background}\label{background}
In this section, we provide background information required for the working principals of the Algorand blockchain, \gls{defi} activity on it with a focus on \gls{dex} protocols, and an overview of \glspl{amm}.

\subsection{Algorand Blockchain}

For the detailed analysis and evaluation of \gls{mev} searching in \gls{fcfs}, our focus is on the Algorand blockchain~\cite{chen2017algorand, cryptoeprint:2017/454, 10.1145/3132747.3132757}, introduced by Silvio Micali in 2017.
Algorand utilizes a consensus mechanism called \gls{ba}, providing instant finality, scalability (in the number of nodes and transactions per second), and avoiding soft forks~\cite{chen2017algorand,cryptoeprint:2017/454,10.1145/3132747.3132757}. 
It employs the \gls{ppos} for Sybil resistance, allowing anyone with at least one ALGO (the native token of Algorand) to join the consensus. 
Unlike Ethereum, Algorand does not reward consensus participants with fixed block rewards or transaction fees. However, discussions about changing incentives are ongoing\footnote{\url{https://github.com/algorand/go-algorand/pull/5740}}.

Regarding high-level specifications, the system can handle around 8000 transactions per second and publishes blocks every \SI{3.3}{\second} in the v3.19.0 release\footnote{\url{https://github.com/algorand/go-algorand/releases/tag/v3.19.0-stable}}. The network consists of approximately \SI{1100}{} nodes (relay and participation nodes)\footnote{\url{https://metrics.algorand.org/}}. Participation nodes connect via relays, each connected to four randomly selected relays. The relays, in turn, forward messages to four relays and all incoming peers. Default connections on a relay are set to \SI{2400}{}. Configuration parameters may vary for each peer, as they are not enforced. Regular clients not participating in consensus also rely on connections via relays, but the number of such clients is unknown. It is expected that in the next year, Algorand will switch from the relay structure to a gossip-based \gls{p2p} layer similar to Ethereum\footnote{\url{https://github.com/algorand/go-algorand/issues/5603}}.

Algorand handles scaling in the number of consensus participants by selecting a subcommittee from the total number of active participation nodes. One consensus round involves block proposer selection, soft vote, and certify vote, with a new committee selected at each step. Committees have varying sizes with \SI{20}{} block proposers, \SI{2990}{} parties in the soft vote, and \SI{1550}{} in the certify vote~\footnote{\url{https://github.com/algorandfoundation/specs/blob/master/dev/abft.md}}. The likelihood of selection correlates with stake amount, and members are unknown until they cast votes. The proposer selection step is crucial for \gls{mev} extraction, as the selected proposer's transaction sequence determines the extracted value. Fast connections to relays can aid \gls{mev} searching, but challenges arise with up to 20 proposers being eligible at every round. \gls{fcfs} ordering is not protocol-enforced but comes with the official Algorand node implementation.


Algorand transactions include payments for native token transfers, \gls{asa} transfers for other token types, and \gls{asc} application calls, which can be split up into \SI{256}{} inner transactions depending on the application complexity. While assets are managed by \gls{asa}  transactions, \gls{asc} applications deploy functions on the Layer-1 network that are written in \gls{teal}, an assembly-like language interpreted in the \gls{avm}, with each having a unique ID. Transaction cost is a fixed minimum of \SI{0.001}{ALGO} per transaction, charged only for successfully executed transactions. The fee strategy shifts to a dynamic cost model per byte during network congestion, which every node determines for itself based on the number of transactions in its local mempool—leading to increased fees if a node is congested. Using multiple clients is crucial to gauge network capacity and adapt fees when congestion occurs over the whole network.

\subsection{Decentralized Finance}
Financial applications built on blockchains, often referred to as Decentralized Finance (DeFi), represent an emerging collection of applications that emulate services found in the traditional finance sector. By employing smart contracts, these applications on blockchains encompass a wide array of services, ranging from decentralized exchanges and options markets to lending protocols and tokenized assets. According to DefiLlama~\cite{noauthor_defillama_nodate}, as of December 2023, Algorand's \gls{defi} ecosystem holds a \gls{tvl} of \SI{82}{million} USD. This valuation has shown stability on an annual basis but exhibits growth over a longer period. Key indicators of blockchain activity include daily volume and the number of daily active addresses. Over the past year, the average daily volume was \SI{5000000} ALGO, with around \SI{36000} daily active addresses, both metrics maintaining consistency year-over-year.
In the realm of Algorand's \gls{dex} ecosystem, Tinyman\footnote{\url{https://tinyman.org}} and Pact\footnote{\url{https://www.pact.fi}} each reported a TVL of \SI{15}{million} USD. HumbleSwap\footnote{\url{https://www.humble.sh/}}, though smaller, also made a significant impact with a \gls{tvl} of \SI{3}{million} USD. Focusing on transaction volumes for December 2023, Tinyman (versions 1 and 2) showed an average daily volume of \SI{729000} USD, followed by Pact with \SI{418000} USD, and HumbleSwap at \SI{41000} USD~\cite{noauthor_defillama_nodate}.

\subsection{Automated Market Makers}
Conventional centralized cryptocurrency exchanges process over \SI{80}{billion} daily in spot-volume\footnote{\url{https://www.coingecko.com/en/exchanges}}, using \gls{clob} architecture. In \glspl{clob}, a centralized entity controls customer asset custody and trade settlement, maintaining an enumerated order list for real-time buyer-seller matching. Orders are processed sequentially, with the central operator acting as an intermediary. In contrast, \glspl{dex} usually utilize smart contracts and an \gls{amm} model instead of on-chain order books due to their simplicity and efficiency in order matching without an intermediary, even in illiquid markets. \glspl{amm} manage liquidity pools, balancing various token reserves, while facilitating \gls{p2p} trading at preset rates determined by mathematical formulas called swap invariants. \gls{cpmm}, a common swap invariant variant used by popular \glspl{dex} such as Uniswap\footnote{\url{https://uniswap.org}} on Ethereum and Tinyman, ensures the product of the assets reserves' in the liquidity pool remains constant unless there is liquidity provision or withdrawal, enabling automated price discovery. Users trade tokens with \glspl{amm}, paying a fee to liquidity providers (usually  0.3\%) and receiving purchased tokens from the pool's reserves. When trading on \glspl{dex}, price slippage refers to the difference between the expected and actual execution prices of the trade. Expected price slippage is the predicted price change based on a trade's volume and \gls{amm}'s liquidity and swap invariant. However, as expected slippage is calculated using historical blockchain data, fluctuations between transaction submission and execution can lead to unexpected slippage. Together, expected and unexpected slippages determine a trade's overall market impact.

\section{Related Work}
As our methodology is focused on profitable opportunity discovery and \gls{mev} extraction on \gls{fcfs} networks, we provide an overview of existing work on these subjects.


\subsection{Profitable Opportunity Discovery}
The discourse on profit-yielding opportunity discovery on blockchain networks includes a range of works focusing on cyclic arbitrage discovery~\cite{zhou_just--time_2021, wang_cyclic_2022, McLaughlin_arb}, even with optimal routing~\cite{angeris_optimal_2022}, sandwiching~\cite{sando}, or more generalized strategies like imitation attacks~\cite{imitation}. Zhou et al.~\cite{zhou_just--time_2021} provide an approach utilizing a greedy cycle detection algorithm paired with a gradual increment-based input searching. While they achieve sub-block time runtime, the algorithm is not optimized for efficiently finding an input and maximizing discovered profits as it operates on a limited asset set. Wang et al.~\cite{wang_cyclic_2022} focus solely on constant-product markets on UniswapV2\cite{UniswapOriginal} and show the consistent existence of greater than 1 ETH opportunities across blocks. Following, McLaughlin et al.~\cite{McLaughlin_arb} conduct a larger scale study and incorporate further market invariants such as UniswapV3~\cite{UniswapV3}, which enables liquidity providers to bound liquidity to a tick range instead of the default full price range as in V2. Their arbitrage discovery study on \SI{5.5}{million} blocks outputs approximately \SI{4.5}{billion} potential opportunities. However, they also show that, out of \SI{20.6}{million} arbitrages which yield over \SI{1}{ETH} revenue, only \SI{0.51}{\%} of them are successfully executable. In contrast, close to \SI{97}{\%} fail due to reverting or incompatible tokens.

\subsection{MEV on FCFS Networks}
Carillo and Hu~\cite{carrillo_mev_2023} analyze arbitrage \gls{mev} extraction on Terra Classic, an \gls{fcfs} blockchain with fixed gas prices. Their study identifies more than \SI{188000} arbitrages, and highlights the success of \gls{mev} searchers adopting a spamming strategy, involving sending multiple failed transactions for every successful one, to gain latency advantage. They also demonstrate the significance of a searcher's geographical position in the network for reducing transaction propagation latency and observing an opportunity generating transaction before the others by monitoring \SI{400000}{} transactions through their \SI{84}{} nodes distributed to different locations. 

Öz et al. in~\cite{oz_study_2023} investigate the applicability of transaction ordering techniques observed in fee-based blockchains to \gls{fcfs} blockchains with a focus on Algorand. Their empirical research, detecting \SI{1.1}{million} arbitrages across \SI{16.5}{million} blocks, sheds light on \gls{mev} extraction approaches under the impact of the \gls{fcfs} network. Their analysis results show the prevalence of network state backruns among all detected arbitrages, with a uniform distribution to block positions. However, they also reveal that particular \gls{mev} searchers profit from top-of-the-block positions, hinting at latency and transaction issuance timing optimizations. Although their on-chain data analysis does not offer new insights into latency games played between searchers and Algorand block proposers, they highlight the use of network clogging as a viable strategy in \gls{fcfs} networks due to its low cost and show the existence of arbitrage clogs on Algorand.

\section{MEV Discovery}\label{disco}
The initial stage of our methodology involves identifying profitable opportunities on Algorand's \gls{fcfs} network. Given the substantial number of arbitrages on Algorand~\cite{oz_study_2023}, we focus on developing an algorithm to detect these opportunities. Prior research on Ethereum by Zhou et al. ~\cite{zhou_just--time_2021} demonstrated real-time cyclic arbitrage detection using a greedy method on a small set of assets, with trade input discovery through gradual increments. In this work, we propose a real-time algorithm tailored to Algorand's specific time constraints, aiming to identify nearly all emerging arbitrage cycles and incorporate a more efficient input optimization technique.


To evaluate our algorithm's efficacy, we collect state data from Algorand and execute the algorithm subsequent to each block's finality. We present results in both unconstrained and time-constrained settings, with the latter being more relevant for the competitive dynamics of an \gls{fcfs} network. This approach allows us to assess the algorithm's practicality and efficiency in a real-world blockchain environment.


\subsection{Cyclic Arbitrage Detection Algorithm}\label{arbie}
Our methodology for detecting profitable arbitrage opportunities begins with a setup phase, where we translate the space of assets \(A\) and pools \(P\) into a multigraph \( G = (V, E) \). Here, \( V \) represents the set of vertices, each corresponding to an asset \( a \in A \), and \( E \) is the set of edges, where each edge denotes a pool \(p_{ij} \in P\) that enables the exchange between assets \(a_i\) and \(a_j\)\footnote{For the scope of this study, we ignore the pools which offer more than two assets.}. Given that \( G \) is a multigraph, we define \( E_{ij} \subseteq E \) as the set of pools available for exchanging the two assets.

Assuming a subset $PA\subseteq A$ as the profit assets of interest, we employ a cycle detection algorithm on $G$ to find the set of cycles $C$ of length $l\in L$ for each profit asset $pa\in PA$, denoted as $C^l_{pa}$. Each trading cycle $c^l_{pa} \in C^l_{pa}$ comprises $l$ swaps $\{s^1, \ldots, s^l\}$ where the input of $s^1$ and the output of $s^l$ is the profit asset $pa$. Each swap $s_{ij}$ between assets $a_i$ and $a_j$ can be implemented in $|E_{ij}|$ ways. Therefore, a cycle $c^l_{pa}$ can have $N_{c^l_{pa}}$ different implementations, where \(N_{c^l_{pa}}=\prod_{i=1}^{l}|E_{s^i}| \) and \( I_{c^l_{pa}} \) represents the set of implementations, with \( |I_{c^l_{pa}}| = N_{c^l_{pa}} \). The aggregate set of implementations for cycles of length $l$ for a profit asset \( pa \), denoted as \( I^l_{pa} \), is defined as \( I^l_{pa} = \bigcup_{c^l_{pa} \in C^l_{pa}} I_{c^l_{pa}} \). The comprehensive implementation set for all assets in \( PA \) is \( I_{\text{total}} = \bigcup_{{pa \in PA},{l \in L}} I^l_{pa} \).

The setup stage outputs \( I_{\text{total}} \), leading to the arbitrage detection phase, which happens in real time after every proposed block \( b \). In this phase, we apply the detection algorithm to a subset of cycle implementations \( I^b_{\text{total}} \subseteq I_{\text{total}} \) related to our profit assets, specifically targeting cycles with updated pool reserves in block $b$. We ignore the cycles with only stale pools as we already evaluate the opportunities on them previously.

The algorithm is confined to operate within a predefined time window \( \tau \), which, for \gls{fcfs} networks, depends on the arrival time of the first transaction, changing a relevant pool's state. In such networks, the desired position in a block can only be achieved by correctly timing the transaction issuance and propagation to the network. In fee-based blockchains such as Ethereum, $\min \tau$ is equal to block time (ignoring network propagation latency), as the targeted position can still be obtained by issuing a transaction with sufficient fees at any time before the block is mined.

While \(\tau\) is not reached, for each cycle \( i \in I^b_{\text{total}} \), we check if the product of involved pools' exchange rates' is greater than one, indicating an arbitrage opportunity. If so, we search for the profit-maximizing input for $i$ using SciPy's\footnote{https://scipy.org/} \textit{minimize function} constrained by the involved pools' swap invariant. This approach maximizes our profitability objective function in a constraint-free nonlinear optimization landscape, employing solvers adept in numerical gradient approximation. In case the profit level of the arbitrage is more significant than a lower limit, we append it to the set of candidate arbitrages for block $b$, denoted as \( \mathcal{A}^b_{\text{candidates}} \).

The final stage of our approach involves one of two strategies for arbitrage selection: a greedy strategy for maximizing profits (which we assumed as default in the rest of the analyses) and an \gls{fcfs} strategy. The former strategy initially iterates over every candidate arbitrage in \( \mathcal{A}^b_{\text{candidates}} \) and, after evaluating all, greedily selects and issues the most profitable, non-overlapping set, which we denote with $\mathcal{A}^b_{\text{greedy}}$. On the other hand, the \gls{fcfs} strategy does not wait to check the profitability of every arbitrage in \( \mathcal{A}^b_{\text{candidates}} \). It issues them as soon as their optimal input is discovered. While this strategy ensures rapid execution, it trades off maximizing profitability since early discovered arbitrages may invalidate to-be-realized, more lucrative opportunities. Although there is no deliberate selection of arbitrages in the \gls{fcfs} strategy, we use $\mathcal{A}^b_{\text{FCFS}}$ to denote the set of arbitrages that will be executed without getting invalidated by any prior arbitrage. Every arbitrage in $\mathcal{A}^b_{\text{greedy}}$ and $\mathcal{A}^b_{\text{FCFS}}$ interacts with a unique set of pools, as otherwise, one arbitrage could invalidate another one by changing the reserves of the intersecting pool.

\subsection{Empirical Evaluation Setup}
To test the performance of our algorithm, we constructed a historical state data collection setup. Leveraging the API capabilities of the Algorand node, we establish a process that continuously listens to our node by invoking the \texttt{/v2/status/wait-for-block-after/round} endpoint\footnote{\url{https://github.com/algorand/go-algorand/algod/api/server/v2/}}. This method sets up an internal wait channel within the node that unblocks upon reaching the desired round, thus notifying us of new blocks. Concurrently, we utilize SDK utilities of various \glspl{dex} on the Algorand network to fetch reserve data of pools following the CPMM price invariant. This data, essential for arbitrage discovery, is stored for each round, maintaining a continuous and comprehensive market overview. ALGO token price data for revenue calculation is fetched from \cite{gecko}.

\subsection{Limitations}
In our methodology, we faced certain constraints that are important to consider. Firstly, there is no known forking tool on Algorand (such as Ganache\footnote{\url{https://github.com/trufflesuite/ganache}} on Ethereum) which would allow us to construct the blockchain state at a desired block and execute a transaction. Thus, we built our pipeline for constructing historical states. Due to the limitation of this approach, we focused on only a subset of assets, pools, and \gls{dex} protocols. For example, we had to exclude the HumbleSwap \gls{dex} as it did not provide an SDK with low latency. Hence, the arbitrages we discover on the finalized blocks only represent a lower bound.
Additionally, our approach focused on collecting state data on finalized blocks rather than a more comprehensive network-level data collection on the mempool, likely leads to underestimating total arbitrage opportunities. We also simplified our financial modeling, disregarding flash loan fees and assuming the immediate availability of initial assets, which might not reflect real-world trading conditions. 

\subsection{Results}
We have tracked \SI{136} assets, exchanged in \SI{255}{} pools, on three different \glspl{dex} (TinymanV1, TinymanV2, Pact), starting from block \SI{32608011}{} (Thu, 05 Oct 2023 00:49:42 GMT) until block \SI{33039007}{} (Sat, 21 Oct 2023 21:05:40 GMT). In these 16 days, \SI{430996}{} blocks were built on the Algorand blockchain. In \SI{30828}{} (\SI{7.1}{\%}) of them, reserves of at least one pool we tracked got updated, and we ran the arbitrage detection algorithm on it.

\subsubsection{Unconstrained Arbitrage Discovery}\label{unconstrained}
Before analyzing the time-constrained performance of our algorithm, we let it run unconstrainedly to observe the maximum profitability of block state arbitrages. To benchmark its performance, we also calculate the executed arbitrage profits in the same block range, utilizing the heuristics defined in \cite{oz_study_2023}.

\begin{figure}[t!]
\centering
\includegraphics[width=0.5\linewidth]{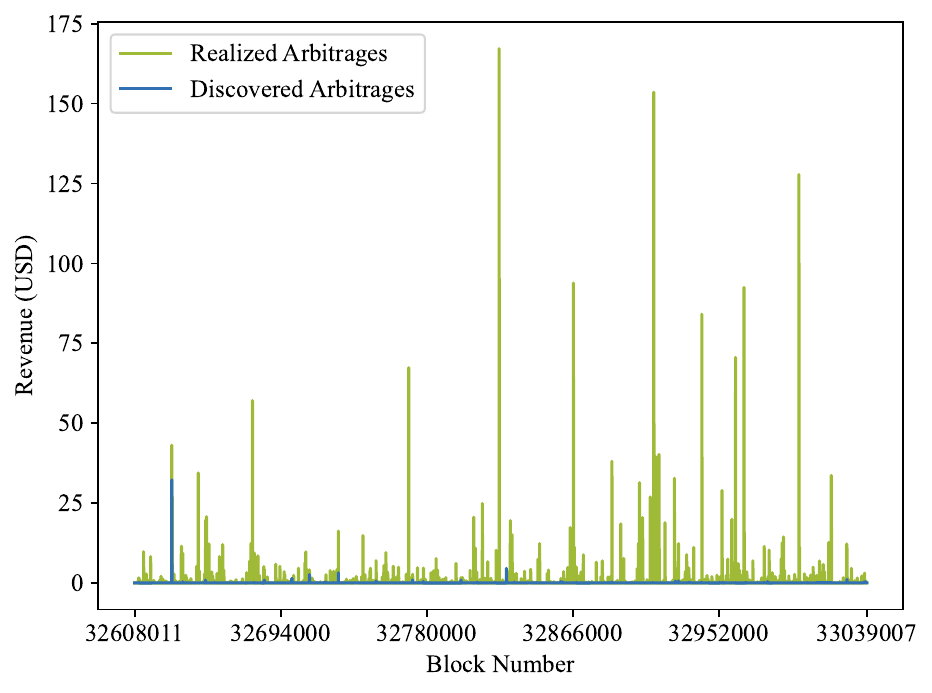}
\caption{Time series of arbitrage revenue across the blocks in range \SI{32608011}{} to \SI{33039007}{}. The blue plot reflects the revenue discovered on the block state by our algorithm, while the green plot shows the total realized revenue by arbitrages executed in every block.}
\label{fig:unconstrained}
\vspace{-2mm}
\end{figure}

\Cref{fig:unconstrained} displays the revenue plots for arbitrages discovered through our algorithm, which we only ran on finalized block states (blue) versus the arbitrages executed in reality (green). The stark dominance of realized arbitrage revenues showcases that \gls{mev} searchers on Algorand promptly exploit arbitrage opportunities inside the block they emerge. Hence, in most cases, price discrepancies do not carry over to the next block. While the maximum realized revenue of an arbitrageur is \SI{167.17}{USD}, we find, at most, a \SI{32.2}{USD} opportunity on the block state that is fully closed in the subsequent six blocks. When we manually checked the most profitable ten arbitrages for the window between the position of the opportunity-creating transaction and their respective backruns, we found that the first backrun was always located at the immediate following position.

The efficiency of \gls{mev} searchers on Algorand in arbitrage execution, in line with the results from~\cite{oz_study_2023}, showcase that our initial attempt for searching \gls{mev} only at the block state level is a naive approach. In future work, we plan to collect transaction data on the mempool directly and execute our algorithm after every trade on a pool relevant to us. This way, we can detect the opportunities when they appear during block construction (not only after the block is finalized) and find their optimal revenue. Currently, we cannot fully assess whether \gls{mev} searchers backrunning on the network-level run their strategies with optimized inputs.


\subsubsection{Time-Constrained Arbitrage Discovery}
The success of an arbitrage strategy depends on execution prior to an update in the reserves of the arbitraged pools. We have introduced $\tau$ to denote the time window before such an update occurs as part of a competing arbitrageur's transaction or by an innocent user trade. A competitive arbitrage discovery strategy needs to operate under $\tau$. Hence, in this section, we measure our algorithm's performance with respect to a spectrum of $\tau$ values.

\begin{figure}[t!]
\centering
\includegraphics[width=0.5\linewidth]{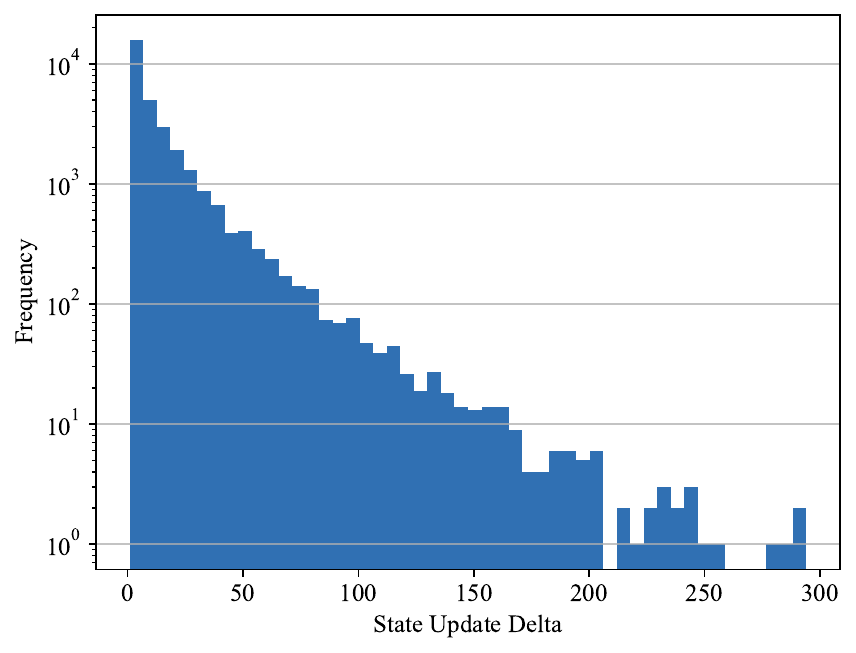}
\caption{Histogram of state update deltas among \SI{430996}{} blocks built between Thu, 05 Oct 2023 and Sat, 21 Oct 2023.}
\label{fig:state_deltas}
\vspace{-2mm}
\end{figure}

Our initial experiments on \SI{430996}{} Algorand blocks, in which only \SI{7.1}{\%} of them have an updated pool we track, show that pools relevant for our arbitrage detection algorithm are updated on median every six blocks (\SI{25}{\%}: \SI{2.0}{}, \SI{75}{\%}: \SI{17.0}{}); hence \( \tau \) is close to \SI{19.8}{\second} (block time $\approx$ \SI{3.3}{\second}). \Cref{fig:state_deltas} displays the distribution of state update deltas for the examined date range. Interestingly, the max state update delta reaches \SI{294}{} blocks ($\sim$\SI{16}{\minute}), although our algorithm does not detect any profitable block state arbitrage in this window.

\paragraph{The Value of Time}
We measure the profitability of our algorithm as a function of $\tau$ to observe the impact of the available runtime window on the discovered value. Although we have detected a median state update delta of six blocks ($\tau = 19.8$),  due to the competitive nature of intra-block opportunities (see \Cref{unconstrained}), we conduct experiments on a range of $\tau \in [0.2, 19.8]$. Additionally, we consider $\tau=\infty$ to encapsulate the maximum profitability in that block.

\begin{figure}[b!]
\centering
\includegraphics[width=0.5\linewidth]{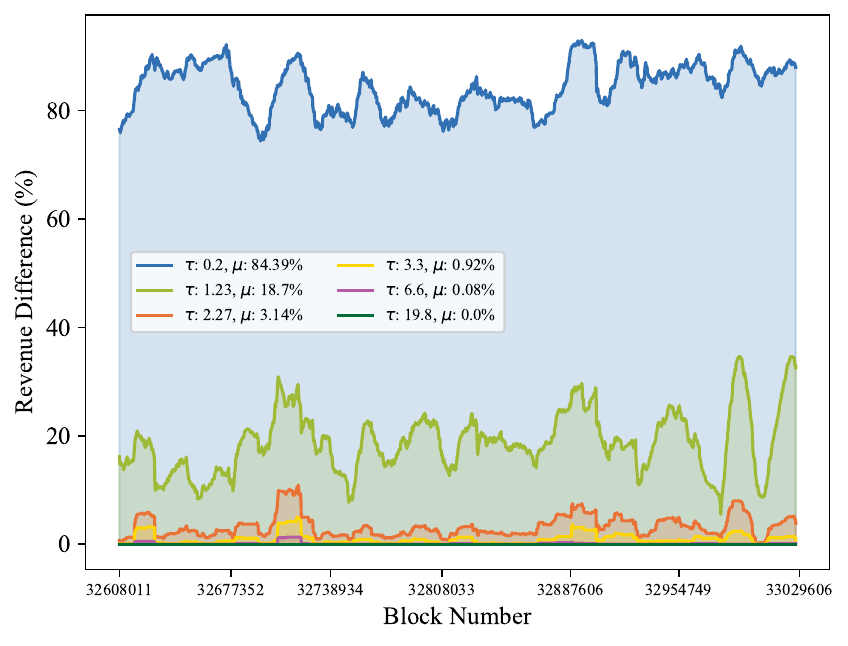}
\caption{Discovered revenue difference (in \%) between $\tau$ values and $\tau=\infty$ over time.}
\label{fig:tau_inf}
\vspace{-2mm}
\end{figure}

\Cref{fig:tau_inf} displays the revenue difference percent of $\tau$ values to maximum discoverable revenue when $\tau=\infty$, with the mean difference ($\mu$) indicated in the legend of the plot. The results indicate that the discovered arbitrage revenue only significantly degrades when $\tau$ is very low (at \SI{0.2} {\second}, \SI{84.39}{\%} less arbitrage revenue is found). On the other hand, almost maximum profitability is reached when $\tau$ is close to block time (\SI{3.3}{\second}). While the revenue difference we observe depends on the infrastructure we execute the algorithm to measure the runtimes and the size of the pool set we consider, our analysis yields an intuition about the positive influence of available runtime on the discovered value.

\begin{figure}[t!]
\centering
\includegraphics[width=0.5\linewidth]{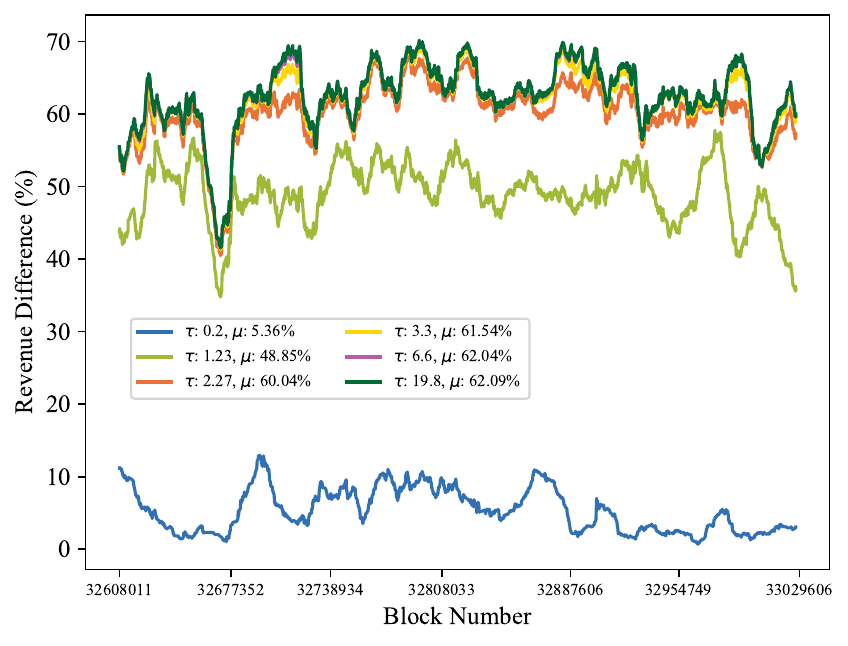}
\caption{Discovered revenue difference (in \%) between profit-maximizing arbitrage selection and FCFS selection strategies, measured for varying $\tau$ values over time.}
\label{fig:max_fcft}
\vspace{-2mm}
\end{figure}

\paragraph{First-Come-First-Served}\label{fcfs}
So far, we have adopted the profit-maximizing, greedy arbitrage selection strategy, as discussed in \Cref{arbie}. However, since we have observed that \gls{mev} searchers on Algorand do not leave a significant amount of profits for block state arbitrage, we need to optimize the runtime of our algorithm further to be competitive on the network-level arbitrage. To that extent, we adopt an \gls{fcfs} strategy for arbitrage selection (as presented in~\cref{arbie}), which does not wait to consider all arbitrages available and select the most profitable ones but issues them as soon as their optimal input is calculated. While this strategy can yield less optimal revenue, it potentially saves valuable time.
\newpage
\Cref{fig:max_fcft} displays the revenue difference between \gls{fcfs} and profit maximizing strategies, for different $\tau$ values. While the disparity is only \SI{5.36}{\%} when the algorithms are run for \SI{0.2}{\second}, the difference between the two strategies becomes more significant with increasing $\tau$ values. This is because, with more time, the profit-maximizing strategy discovers a wider set of opportunities and considers all of them when choosing the most profitable arbitrages to be issued. \gls{fcfs} strategy, on the other hand, will always execute the first arbitrage it finds; hence, with increased time, the probability of finding an arbitrage that overlaps with an already taken one also increases. To minimize the revenue difference between the two approaches, the \gls{fcfs} strategy requires applying a prioritization rule on the candidate arbitrage cycles before processing them. In our experiments, we sorted candidate cycles based on existing liquidity in involved pools, although more sophisticated rules can be developed by modeling the problem in a machine-learning domain.

\section{MEV Extraction}
In the \gls{mev} discovery stage (see \Cref{disco}), we have shown that although the pools we track get updated on median every six blocks, our algorithm fails to find many profitable opportunities as extracted value shows the efficiency of \gls{mev} searchers on Algorand in exploiting arbitrages as soon as they appear in a block. To be competitive in network-level \gls{mev} searching, we need to frontrun other searchers in execution. Since \gls{fcfs} \gls{mev} extraction depends on \textit{latency optimization}, in this section, we demonstrate how transactions can be prioritized by running experiments on a privately deployed Algorand network.

\subsection{Network Experiments and Analysis}\label{sec:network_experiments}
This section introduces the conducted network experiments, primarily focusing on the Algorand transaction ordering mechanism under simultaneous transaction attempts by competing entities. 
The central hypothesis driving these experiments examines the potential for a competitor to secure a transaction position in a block, thereby extracting arbitrage opportunities before others. The arbitrage trader must place their transaction ahead of the competition in a block, as subsequent transactions that compete for the same opportunity are likely to fail once the opportunity has been extracted.  To understand the behavior of the network in practice, we run a sequence of experiments in a private network that helps us verify the behavior on Algorand's mainnet.
The objectives of these network experiments are threefold:

\begin{enumerate}[label={\bfseries Q\arabic*}, leftmargin=0.80cm]
\item How do the latency and transaction fees affect the transaction ordering when two different entities compete for prior transaction execution? 
\item How does the connectivity between relay and participation nodes with differently distributed stakes affect the transaction ordering?
\item Is it possible to prioritize a transaction if we have visibility over the network topology and if so, how can we achieve this?
\end{enumerate}

We set up a private network similar to Algorand mainnet to answer these questions instead of utilizing Algorand's testnet. Even though running experiments on the testnet would also have been costless, we require more visibility and control over the distribution of participation nodes and their stake than in the testnet.

A private network provides granular control over the topology, in which we can introduce various latencies, distribute stakes among the participation nodes the way we want, and scale transaction issuance to emulate various network conditions. 
We use the METHODA framework~\cite{rezabek2023multilayer} that extends the EnGINE toolchain~\cite{rezabek2022engine}. METHODA also implements the Algorand blockchain, making it easier to automate the Algorand network deployment process in a distributed environment relevant to our scenarios. In addition, it supports scalable experiments with a number of nodes and can emulate delays using \gls{netem}\footnote{\url{https://man7.org/linux/man-pages/man8/tc-netem.8.html}}.

\subsubsection{Experiment Design}
\begin{figure*}[t!]
\centering
  \begin{subfigure}[b]{0.3\textwidth}
  \centering
    \includegraphics[width=\linewidth]{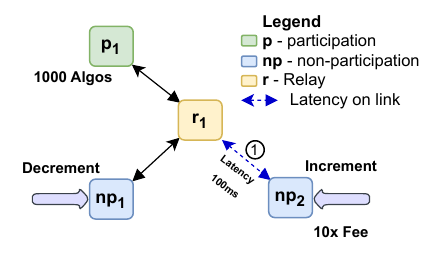}
    \caption{Scenario 1}
    \label{fig:scenario1}
  \end{subfigure}
  \hfill
  \begin{subfigure}[b]{0.3\textwidth}
  \centering
    \includegraphics[width=\linewidth]{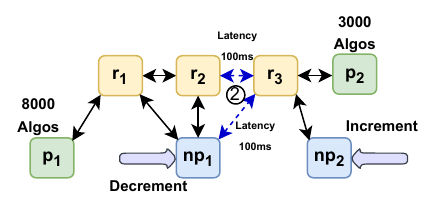}
    \caption{Scenario 2}
    \label{fig:scenario2}
  \end{subfigure}
  \hfill
  \begin{subfigure}[b]{0.3\textwidth}
  \centering
    \includegraphics[width=\linewidth]{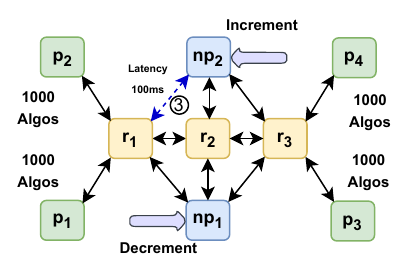}
    \caption{Scenario 3}
    \label{fig:scenario3}
  \end{subfigure}
\caption{Network topology of three scenarios discussed in Section \ref{sec:network_experiments} with legend on in Figure a). Dashed blue arrows connecting peers corresponds to delay of \SI{100}{\ms}.}
\label{fig:topologies}
\end{figure*}

To collect insights and answer the defined questions, we devise three scenarios. These scenarios are realized using three distinct network topologies, each denoted using specific notations. These notations are defined as follows:

\begin{itemize}
    \item \( P = \{ p_i \mid i = 1, \ldots, n_P \} \): Represents the set of participation nodes, where each \( p_i \) is an individual participation node and \( n_P \) is the total number of such nodes.
    \item \( R = \{ r_j \mid j = 1, \ldots, n_R \} \): Denotes the set of relay nodes, with each \( r_j \) being a relay node and \( n_R \) the total number of relay nodes.
    \item \( NP = \{ np_k \mid k = 1, \ldots, n_{NP} \} \): The set of non-participation nodes, where \( np_k \) is a non-participation node and \( n_{NP} \) is their total count.
\end{itemize}
The experimented network topologies are then formally represented as tuples \( (P, R, NP) \), encompassing the node sets and their connectivity relations. The three scenarios we tested as part of this research work have been elaborated in~\Cref{fig:topologies}. \sceni~in ~\Cref{fig:scenario1} introduces a topology with two non-participating peers and a single relay connected to a single participation node. We note that $np_2$ has worse delay towards $r_1$ (\encircled{1}). Therefore, we expect transactions issued through $np_1$ to be positioned in prior block positions even though $np_2$ offers more transaction fees. Similarly, in \scenii,~we introduce a delay (\encircled{2}) from $np_1$ towards $r_3$ and between $r_2$ and $r_3$ (see ~\Cref{fig:scenario2}). However, since the participation node with a higher stake ($p_1$) is expected to propose blocks more often than $p_2$, $np_2$ should be in a worse position regarding transaction prioritization. Lastly, \sceniii~ assigns the same stake to every proposer while placing a delay on the shortest path towards $p_1$ and $p_2$ (\encircled{3}) from $np_2$ (see ~\Cref{fig:scenario3}). As all participation nodes have equal stakes, they have the same probability of winning the proposer selection round, with the only difference being the \gls{vrf} value they generate. Thus, we expect $np_1$'s transactions to be prioritized significantly in half of the experiments whenever $p_1$ or $p_2$ proposes a block.

To emulate two \gls{mev} searchers $S_1$ and $S_2$, simultaneously attempting to exploit an arbitrage opportunity, using the same detection algorithm, we deploy a simple smart contract (see~\Cref{alg}) which tracks the state of a global variable $last\_executed$, and the winner searcher is the one which modifies this variable's value the first. We assume that $S_1$ and $S_2$ change the global state by calling functions \textit{decrement} and \textit{increment}, respectively. 


\begin{algorithm}
\caption{Smart Contract Operations}
\label{alg}
\begin{algorithmic}[1]
\State \textbf{Initialize} global variable $last\_executed = ""$
\Function{increment}{}
    \State $last\_executed \gets "increment"$
\EndFunction
\Function{decrement}{}
    \State $last\_executed \gets "decrement"$
\EndFunction
\end{algorithmic}
\end{algorithm}

\subsubsection{Experiment Setup}
Each Algorand node - participant, non-participant, and relay node is deployed using a combination of multiple physical machines and \glspl{lxc} interacting with each other. Each peer node runs in its own \gls{lxc} container allocated the recommended HW specs - 16 vCPUs for a relay and 8 vCPUs for participation nodes. The \gls{lxc} containers run on a node with 32cores/64threads AMD EPYC 7543 with \SI{512}{GB} DDR 5 RAM. Depending on the scenario, we scale to additional physical nodes once we reach the available resources. The nodes are interconnected via a \SI{10}{GbE} switch. Therefore, even if the delay between the container and physical machine differs, it is negligible compared to introducing delay by netem. We conducted our experiments using the Algorand protocol version \textit{abd3d4823c6f77349fc04c3af7b1e99fe4df699f} and \texttt{go-algorand} in 3.18.1 release. During each experiment, we create 500 blocks corresponding to roughly \SI{27.5}{\minute} of runtime and simultaneously issue transactions from $S_1$ and $S_2$ at every block.

\subsubsection{Experiment Analysis}
To highlight the findings from the three scenarios, we monitor how often the \textit{decrement} method is invoked on node $np_{1}$ by $S_1$, while the \textit{increment} method is simultaneously invoked on node $np_{2}$ by $S_2$. A detailed analysis of these results is provided below, with all results summarized in \Cref{tab:scenarios-1-2-3} highlighting the proposer node, block count, the method, and frequency for the corresponding method execution.

\paragraph{Scenario 1}
In \sceni, we observe a consistent prioritization pattern throughout all \SI{500} iterations. The \textit{decrement} function call directed to node $np_{1}$ by $S_1$ always reaches $p_1$ first over the \textit{increment} call sent to node $np_{2}$ by $S_2$. Notably, this occurs despite the transaction to $np_{2}$ being attached with a fee tenfold that of the transaction to $np_{1}$. This outcome distinctly indicates that higher transaction fees are ineffective within Algorand's \gls{fcfs} network to secure prioritization (unless a network congestion strategy is adopted as described in~\cite{oz_study_2023}). Furthermore, the absence of additional latency between $np_{1}$ and relay $r_{1}$ results in the transactions of $S_1$ being consistently favored. 

\paragraph{Scenario 2}
Continuing with the observations in \scenii, we see that node $p_{1}$ is chosen as the proposer in \SI{377}{} out of \SI{500}{} ($\sim\SI{75}{\%}$) iterations. Hence, we confirm that proposer selection frequency is positively correlated to the stake in the network ($p_1:\SI{72}{\%}$). Additionally, it is noted that when $p_{1}$ served as the proposer, the \textit{decrement} function call is consistently prioritized. On the other hand, when $p_{2}$ is the proposer, the \textit{increment} call received prioritization, although not every time. Despite node $np_{1}$ having advantageous network positioning due to its connections with all relays, the selection of $p_{2}$ as the proposer enables searcher $S_2$, using node $np_{2}$, to propagate their transactions ahead of $S_1$. This advantage for $S_{2}$ can be attributed to the latency existing between $np_{1}$ and relay $r_{3}$. Nonetheless, as $S_1$ prevails in most cases, the findings signify the importance of a node's connectivity to a relay with higher-staked participants to achieve a prior position in the block. 

\paragraph{Scenario 3}
Lastly, \sceniii~confirms that the proposers' observed selection frequency matches the theoretical selection frequency and its stake within the network. Additionally, two critical observations emerge from this scenario.
First, in instances where $p_{1}$ and $p_{2}$ are selected as proposers, there is a heavy bias towards prioritizing the \textit{decrement} call by $S_1$. This observation validates our earlier conclusions, underscoring latency as a pivotal factor influencing transaction order. The latency between $np_{2}$ and relay $r_{2}$ contributes to the delayed processing of the \textit{increment} function call sent to $np_{2}$ from $S_2$.
Second, when $p_{3}$ and $p_{4}$ are selected as proposers, the data indicate an almost equal likelihood of giving preference to either searcher's call. Hence, this scenario suggests that when searchers' utilized nodes' latency to relays and the respective participation nodes with equivalent stakes are similar, they have an equal chance of executing their competing transaction first.

\input{include/Resultstable}


\subsection{Network Analysis Findings}
The scenarios, emulating critical dynamics on the blockchain network, contribute to verifying Algorand's behavior under various conditions and answer the outlined questions \textit{\textbf{Q1-Q3}}. The summary of the key findings is as follows:

\begin{itemize}
    \item Transaction fee does not play a role in prioritizing a transaction in the Algorand blockchain since the proposers have no incentive to reorder transactions based on fees attached as they do not get to keep them.
    \item Network latency is critical in ordering transactions in \gls{fcfs} blockchains. For arbitrage traders in Algorand, this underscores the importance of minimizing the latency between their transaction nodes and the corresponding relay nodes. Strategic positioning is crucial for enhancing the likelihood of transaction precedence to competitors.
    \item The proximity of a node to a high-staked proposer is significantly influential in transaction sequencing. Arbitrage traders aiming to exploit market opportunities should prioritize transmitting their transactions to well-connected relays with multiple high-staked nodes. Such an approach increases the probability of their transactions being processed earlier, thereby gaining a competitive edge in transaction execution.
\end{itemize}


\section{Conclusion}
This paper provides a two-staged approach for \gls{mev} searching on Algorand, an \gls{fcfs} blockchain. 
Initially, we propose an algorithm for arbitrage detection and evaluate its performance using historical block state data, taking into account the time constraints of Algorand. Our findings indicate that although significant state updates between blocks are infrequent, \gls{mev} searchers on Algorand tend to close arbitrage opportunities within the same block they arise. Therefore, to be competitive, \gls{mev} opportunity detection should occur at the transaction level in the mempool rather than at the block state level as initially attempted. We also show the impact of available algorithm runtime and arbitrage selection strategy on the discovered revenue.

Subsequently, we perform experiments on a private Algorand network to identify key factors in transaction prioritization within an \gls{fcfs} framework. Our results highlight the importance of minimizing latency, particularly in connections to relays with well-connected links to multiple high-staked nodes.

Looking ahead, we plan to refine our arbitrage detection algorithm by considering a broader set of pools and applying it directly at the network level. Additionally, we aim to extend our network experiments to the Algorand mainnet to explore potential latency correlations between high-staked block proposers and successful \gls{mev} searchers. This future research will enhance our understanding of \gls{mev} dynamics in \gls{fcfs} blockchains like Algorand.

\bibliographystyle{plainnat}
\bibliography{sample-bibliography}


\end{document}

%% file: acronyms.tex
\newacronym{bft}{BFT}{Byzantine Fault Tolerant}
\newacronym{ba}{BA}{Algorand Byzantine Fault Tolerance Protocol}
\newacronym{ppos}{PPoS}{Pure Proof-of-Stake}
\newacronym{pow}{PoW}{Proof-of-Work}
\newacronym{mev}{MEV}{Maximal Extractable Value}
\newacronym{nft}{NFT}{non-fungible token}
\newacronym{vrf}{VRF}{Verifiable Random Function}
\newacronym{fp}{FP}{False Positive}
\newacronym{fn}{FN}{False Negative}
\newacronym{dex}{DEX}{Decentralized Exchange}
\newacronym{cex}{CEX}{Centralized Exchange}
\newacronym{fcfs}{FCFS}{First-Come-First-Served}
\newacronym{bti}{BTI}{Batch Transaction Issuance}
\newacronym{amm}{AMM}{Automated Market Maker}
\newacronym{defi}{DeFi}{Decentralized Finance}
\newacronym{asa}{ASA}{Algorand Standard Asset}
\newacronym{asc}{ASC1}{Algorand Smart Contract}
\newacronym{teal}{TEAL}{Transaction Execution Approval Language}
\newacronym{avm}{AVM}{Algorand Virtual Machine}
\newacronym{pga}{PGA}{Priority Gas Auction}
\newacronym{tvl}{TVL}{Total Value Locked}
\newacronym{p2p}{P2P}{Peer-to-Peer}
\newacronym{lxc}{LXC}{Linux container}
\newacronym{netem}{netem}{Network Emulator}
\newacronym{cpmm}{CPMM}{Constant Product Market Maker}
\newacronym{clob}{CLOB}{Central Limit Order Book}

\newcommand{\encircled}[2][0.8mm]{%
    \raisebox{.5pt}{%
        \textcircled{%
            \raisebox{0.35pt}{%
                \kern #1
                \scalebox{0.70}{#2}
            }%
        }%
    }%
}
\definecolor{ourgrey}{rgb}{0.60,0.60,0.60}

\newcommand{\sceni}{\textit{\textbf{Scenario I}}}
\newcommand{\scenii}{\textit{\textbf{Scenario II}}}
\newcommand{\sceniii}{\textit{\textbf{Scenario III}}}

%% file: include/Resultstable.tex
\newcolumntype{C}[1]{>{\centering\arraybackslash}p{#1}}

\begin{table}[hbtp]
\centering
\caption{Summary of Observations for Scenarios 1, 2 and 3}
\label{tab:scenarios-1-2-3}
\begin{tabular}{@{}lcccr@{}}
\toprule
\textbf{} & \textbf{Proposer} & \textbf{Blocks} & \textbf{Prioritized Method} & \textbf{Frequency} \\
\midrule
\multirow{1}{*}{\textbf{Scenario 1}} & $p_{1}$ & 500 & Decrement & 100\% \\
\midrule
\multirow{2}{*}{\textbf{Scenario 2}} & $p_{1}$ & 377 & Decrement & 100\% \\
 & $p_{2}$ & 123 & Increment & 76.42\% \\
\midrule
\multirow{4}{*}{\textbf{Scenario 3}} & $p_{1}$ & 126 & Decrement & 96.03\% \\
 & $p_{2}$ & 113 & Decrement & 97.01\% \\
 & $p_{3}$ & 134 & Increment & 51.97\% \\
 & $p_{4}$ & 127 & Decrement & 54.80\% \\
\bottomrule
\end{tabular}
\end{table}